\documentclass[referee]{raa}            

\usepackage{graphicx,times}             
\usepackage{natbib}
\usepackage{amssymb,amsmath}
\bibpunct{(}{)}{;}{a}{}{,}

\usepackage[pagebackref=true]{hyperref}

\begin{document}

  \title{MCI: Multi-Channel Imager on the Chinese Space Station Survey Telescope
}

   \volnopage{Vol.0 (20xx) No.0, 000--000}      
   \setcounter{page}{1}          

   \author{Zhen-Ya Zheng* 
      \inst{1,12}
   \and Chun Xu 
      \inst{1,12}
   \and Xiaohua Liu 
      \inst{2}
   \and Yong-He Chen
       \inst{2}
   \and Fang Xu 
       \inst{2}
   \and Hu Zhan 
       \inst{3}
    \and Xinfeng Li 
        \inst{4}
   \and Lixin Zheng 
      \inst{1,12}
   \and Huanyuan Shan 
      \inst{1,12}
   \and Jing Zhong 
      \inst{1,12}
   \and Zhaojun Yan 
      \inst{1,12}
   \and Fang-Ting Yuan 
      \inst{1,12}
   \and Chunyan Jiang 
      \inst{1,12}
   \and Xiyan Peng 
      \inst{1,12} 
   \and Wei Chen 
      \inst{1,12} 
    \and Xue Cheng
      \inst{2}
    \and Zhen-Lei Chen 
      \inst{1}
    \and Shuairu Zhu 
      \inst{1}
    \and Lin Long 
      \inst{1}
   \and Xin Zhang 
       \inst{3}
   \and Yan Gong 
       \inst{3}
   \and Li Shao 
       \inst{3}
    \and Wei Wang 
       \inst{3}      
   \and Tianyi Zhang 
       \inst{5}
   \and Guohao Ju 
       \inst{5}
   \and Chenghao Li 
       \inst{5}
   \and Wei Wang 
       \inst{5}
   \and Zhiyuan Li 
       \inst{6}
    \and Tao Wang 
       \inst{6}
   \and Junfeng Wang 
       \inst{7}
   \and Chengyuan Li 
       \inst{8}
   \and Bin Ma 
       \inst{8}
    \and Jianguo Wang 
        \inst{9}
    \and Lei Wang 
        \inst{10}
    \and Dezi Liu 
        \inst{11}
    \and Nie Lin 
        \inst{1,12}
   \and Kexin Li 
      \inst{1,12}       
   \and Xinrong Wen 
      \inst{1,12}
   \and Maochun Wu 
      \inst{1,12}
     \and Ruqiu Lin 
      \inst{1} 
     \and Xiang Ji 
      \inst{1}
   }

   \institute{Shanghai Astronomical Observatory, Chinese Academy of Sciences,
             Shanghai, 200030, China; {\it zhengzy@shao.ac.cn}\\
        \and
             Shanghai Institute of Technical Physics, Chinese Academy of Sciences, Shanghai, 200083, China;\\
        \and
             National Astronomical Observatories, Chinese Academy of Sciences, Beijing, 100101, China;\\
        \and 
            Technology and Engineering Center for Space Utilization, Chinese Academy of Sciences, Beijing, 100094, China;\\
        \and
            Changchun Institute of Optics, Fine Mechanics and Physics, Chinese Academy of Sciences, Changchun, 130033, China;\\
        \and
            School of Astronomy and Space Science, Nanjing University, Nanjing, 210023, China;\\
        \and
            Department of Astronomy, Xiamen University, Xiamen, Fujian 361005, China;\\
        \and
            School of Physics and Astronomy, Sun Yat-Sen University, Zhuhai, 519082, China;\\
        \and
            Yunnan Observatories, Chinese Academy of Sciences, Kunming, 650011, China;\\
        \and
            Purple Mountain Observatory, Chinese Academy of Sciences, Nanjing 210023, China;\\
        \and
            South-Western Institute for Astronomy Research, Yunnan University, Kunming, Yunnan, 650500, China;\\
        \and
            CSST Science Center for the Yangtze River Delta Region, Shanghai, 200030, China.\\
\vs\no
   {\small Received 20xx month day; accepted 20xx month day}}

\abstract{The Multi-Channel Imager (MCI) is a powerful near-ultraviolet (NUV) and visible imager onboard the Chinese Space Station Survey Telescope (CSST). The MCI provides three imaging channels, which are the NUV channel, the Blue channel and the Red channel, with the wavelength range of 255-430 nm, 430-700 nm, and 700-1000 nm, respectively. MCI's three channels can target the same field simultaneously, which is unique compared to other imagers onboard the Hubble Space Telescope (HST) or the James Webb Space Telescope (JWST). Each channel employs a CCD focal plane of 9216 x 9232 pixels and $\sim$7\arcmin.5 x 7\arcmin.5 field of view (FOV), which are about $\gtrsim 4$ times greater than the FOVs of HST imagers.
The MCI's three channels feature unprecedented sensitivities and field of views complement the NUV and visible capabilities of the CSST for high-precision photometry and weak-signal detection, which would help build a new standard-star system and the deepest UV-Optical exposures for CSST. Rich filter sets of MCI would help explore other sciences such as local emission line mapping, high-z Ly$\alpha$ emitters searching, etc. Here we present key design features, results of current ground tests, and suggested observing strategies of the MCI.
}

   \authorrunning{Z.-Y. Zheng, C. Xu, X. Liu, et al. }            
   \titlerunning{Introduction to the CSST-MCI}  

   \maketitle

%
%
\section{Introduction}           
\label{sect:intro}

The Multi-Channel Imager is one of the five first-generation astronomical instruments onboard the Chinese Space Station Survey Telescope (CSST, \citealt{Zhan2011, Zhan2021, CSST2025}). The CSST is a 2-meter space telescope operating at the same orbit of the China Manned Space Station, and is one of the major science projects of the China Manned Space Station Program. It is scheduled to be launched in about 2 years. The five first-generation astronomical instruments include Multi-band Imaging and Slitless Spectroscopy Survey Camera (SC, \citealt{Zhan2021}), THz Spectrometer (TS), Multi-Channel Imager (MCI), Integral Field Spectrograph (IFS), and Cool Planet Imaging Coronagraph (CPI-C). The MCI program was initiated in 2018 after several critical review meetings, aiming to expanding the CSST's capabilities on the high-precision photometry, the weak-signal detection, and related sciences. The MCI instrument is jointly developed by Shanghai Institute of Technical Physics (SITP, CAS) and Shanghai Astronomical Observatory (SHAO, CAS). 

The MCI is designed to divide the optical path into three channels: Near-ultraviolet (NUV, 255-430 nm), Optical-blue (Blue, 430-700 nm) and Optical-red (Red, 700-1000 nm), achieving the function of simultaneous imaging of the three channels. The MCI also employs a relay optical system to extend the focus length, so as to obtain an angular size of 0\arcsec.05 per pixel, which is comparable to that of imagers onboard the Hubble Space Telescope. Each channel of the MCI uses a 9kx9k e2v CCD to cover a 7\arcmin.5$\times$7\arcmin.5 field of view (FOV), which is about $\geq$ 5 times FOV of HST imagers. 
In Tab. \ref{tab:comparison}, we list the information of pixel arrays and the FOV of CSST MCI, in comparison with HST WFC3, HST ACS, HST WFPC2 and JWST NIRCam imagers. We should note that except for MCI, all the other imagers listed in Tab. \ref{tab:comparison} can only do single-channel imaging.

Thanks to the superb sharp and stable optics of the CSST, the MCI is optimized for high-precision photometry and weak-signal detection, on which we set two key programs for the CSST MCI. The MCI's first key program is setting the CSST's photometric standard stars. We will follow the Gaia standard stars program \citep{2012MNRAS.426.1767P, 2021MNRAS.501.2848A} to build the photometric standard stars for the main optical surveys of the CSST. The second key program is the CSST's extremely deep fields (XDFs), which are similar to the HST's famous surveys such as XDF \citep{2013ApJS..209....6I} and Hubble Frontier Fields \citep[HFF,][]{2017ApJ...837...97L}. Currently, we are simulating these MCI key programs \citep{Yan2025, Xie2025}.

\begin{table}
\begin{center}
\caption[]{Comparison of CSST MCI with HST and JWST imagers.}\label{tab:comparison}
 \begin{tabular}{lccccc}
  \hline\noalign{\smallskip}
Channel & Wavelength Range &  Pixel Format & Pixel Scale  &   FOV    & FOV/FOV$_{\textrm{MCI}^*}$  \\
     & (nm)   &      (pixel)         &  (arcsec)    & (arcsec$^2$) &         \\
  \hline\noalign{\smallskip}
{CSST MCI/NUV} & {255-430} & {9216x9232} & {0.050}     & {450x450} & {1.00} \\ 
{CSST MCI/Blue} & {430-700} & {9216x9232 } & {0.050}     & {450x450} & {1.00} \\
{CSST MCI/Red} & {700-1000} & {9216x9232 } & {0.050}     & {450x450} & {1.00} \\\hline %
HST WFC3/UVIS & 200-1000 & 4102x4096 & 0.040     & 162x162 & 0.13 \\
HST WFC3/IR & 850-1700 & 1024x1024 & 0.13     & 123x137  & 0.08 \\
HST ACS/WFC & 350-1050  & 4096x4096 & 0.050     & 202x202 & 0.20 \\
HST WFPC2/WF & 115-1100  & 800x800 (x3) & 0.100     & 80x80 (x3) & 0.09 \\ \hline
JWST NIRCam/NIR & 600-2300 & 4258x4258 (x2)  & 0.031     & 132x132 (x2) & 0.17 \\
JWST NIRCam/MIR & 2400-5000 & 2048x2048 (x2) & 0.063     & 129x129 (x2) & 0.16
 \\ \hline
\end{tabular}
\end{center}
\end{table}

The MCI is equipped with a total of 30 filters to meet the needs of multiple sciences. The filters NUV, u, r, z and y, which are similar to the CSST-SC filters, can be used to assist the flux calibration of the SC. The NUV channel CBU filter, Blue channel CBV filter, and Red channel CBI filter basically cover the spectral range of each channel, and will be used in the observation of the MCI XDF of $\sim$300 arcmin$^2$, with magnitude limits of CBV $\geq$ 30, and CBU and CBI $\geq$ 29. The remaining ones include the narrow, medium and broad filters, mainly used in scientific research for extragalactic celestial objects \citep[e.g.,][]{2022RAA....22b5019C, 2025ApJ...987...94Z, 2022SCPMA..52b9801Y}, galactic objects \citep{2022RAA....22i5004L}, exoplanets \citep{2025ChA&A..49..436D}, and so on.

Here in this work, we present a brief introduction to the CSST-MCI. We introduce the key design features of the instrument in Sec. \ref{sect:Design}, and present some results of the ground tests and the calibration program in Sec. \ref{sect:test}. We suggest some observing strategies of the MCI in Sec. \ref{sect:obs}. Finally, we give a summary in Sec. \ref{sect:sum}. 
Detailed description of the filters, the ground test and calibrations will be presented in separate papers. 

\section{Key Design Features of MCI}
\label{sect:Design}

   \begin{figure}
   \centering
   \includegraphics[width=\textwidth, angle=0]{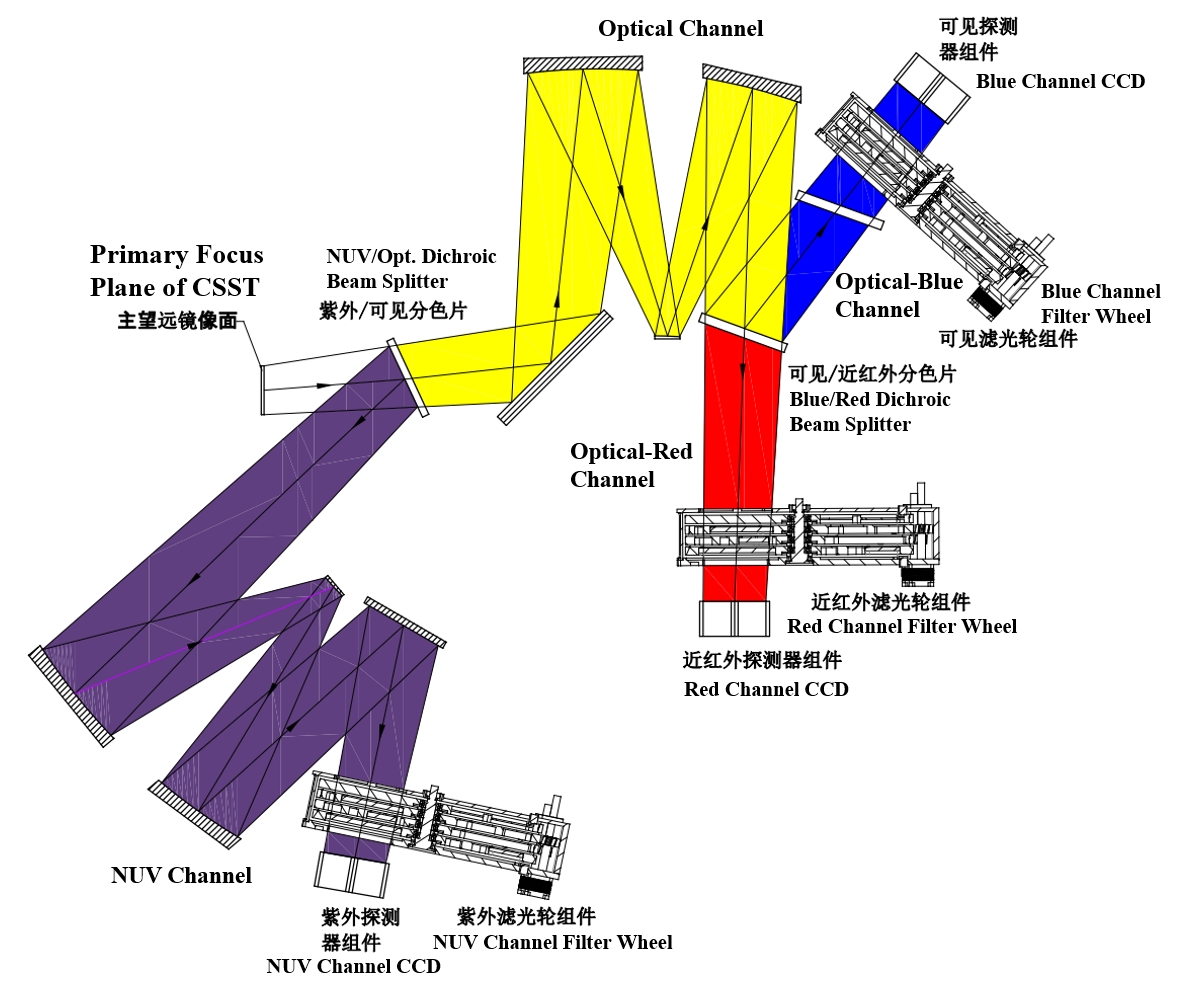}
   \caption{Diagram of MCI components and light path.}
   \label{Fig1}
   \end{figure}
%

In Tab. \ref{Tab1}, we summarize the key scientific specifications for the MCI.
MCI incorporates three channels contiguous from NUV to optical bands: an NUV channel MCI/NUV, an Optical-blue channel MCI/Blue and an Optical-red channel MCI/Red. The short wavelength cutoff depends on the cutoff wavelengths of the coating of the primary mirror and the widest filter CBU. The long wavelength cutoff depends on the quantum efficiency (QE) of the red-sensitive CCD in the MCI/Red channel. In each channel, we have ten filters including ultra-broad, broad, medium and narrowband filters. We choose an e2V CCD for the detector of each channel. The total throughput and sensitivity are estimated in the end of this section.

\begin{table}
\begin{center}
\caption[]{Key scientific specifications for the MCI.}\label{Tab1}


 \begin{tabular}{|c|c|c|c|c|}
  \hline\noalign{\smallskip}
       & NUV Channel &  Blue Channel & Red Channel  &   Units/Notes \\
  \hline\noalign{\smallskip}\hline
Format & 9232 x 9216      &   9232 x 9216     &  9232 x 9216      & pixel \\\hline %
Field & 450 x 450    & 450 x 450      & 450 x 450   &  arcsec$^2$ \\\hline %
Pixel &    0.05      &     0.05       &  0.05       & arcsec \\\hline %
Spectral Range & 255-430 & 430-700 & 700-1000  & nm \\\hline 
Dark Current &  $<$0.02  &  $<$0.02  & $<$0.02   & e$^-$/pixel/sec \\\hline
Readout Noise & $<$5     & $<$5     & $<$5     & e$^-$ per readout \\\hline
Operating Temperature & -100   & -100   & -100      & $^o$C \\\hline
Number of Filters        & 10     & 10      & 10       &   \\\hline
     & CBU & CBV & CBI  & XDF Filters \\
Full-FOV Filters  & F275W, F336W & F606W &   F814W, F850LP & HST Filters \\
  & F373N & F502N, F658N & F815N & Narrow Filters \\\hline
       & NUV, u & r & z, y & SC Filters\\
1/4-FOV Filters & WU & F467M, F555W & F845M, F960M & Broad/Medium Filters \\
     & F280N, F343N, F395N & F487N, F656N, F673N & F925N, F968N & Narrow Filters \\\hline
     {Readout Time} & 40 & 40 & 40 & sec \\\hline
     {Filter-exchange Time} & 210 & 210 & 210 & sec \\\hline
    {Exposure Time} & 1--1200 & 1--1200 & 1-1200 & sec \\
  \hline
\end{tabular}
\end{center}
\end{table}

\subsection{Optical System with Simultaneous Three Channels}

Light from CSST's optical system goes into three different channels, Near-ultraviolet (NUV), Optical-blue (Blue), and Optical-red (Red), achieving simultaneous imaging of the three channels. It uses two relay optical systems to extend the focus and obtain an angular resolution of 0\arcsec.05 per pixel. Each channel of MCI uses an e2v CCD290-99 to cover a 7\arcmin.5 × 7\arcmin.5 FOV, and switches the filter through a rotating wheel to meet the observation needs of different sciences. 
The MCI consists of several key subsystems: relay optics, calibration units, detector assemblies (including cryocoolers), filter wheels, focusing devices, electronic modules, thermal control units, and interface panels.

The diagram of MCI components and light path is shown in Fig. \ref{Fig1}. 
The incoming light from CSST's main optical system passes through the first dichroic beam splitter, where it is divided into two channels, the near-ultraviolet (NUV) and visible light. Each channel employs a relay optical system with three reflecting off-axis mirrors to extend the focal length from 28 meters to 42 meters. Subsequently, the visible light is further split by a second dichroic beam splitter into the optical blue and red light. Therefore, MCI's three channels, NUV, Blue and Red, can work simultaneously. At the end of each channel, the light goes though a specified filter in the filter wheel, received by an e2v 290-99 CCD detector, before which a high-speed shutter is used to control the exposure. A focus adjustment device is set up as well behind the detector.





\subsection{30 Filters in Three Channels}
\label{sect:Design:MCIfilters}

\begin{figure}
   \centering
   \includegraphics[width=\textwidth, angle=0]{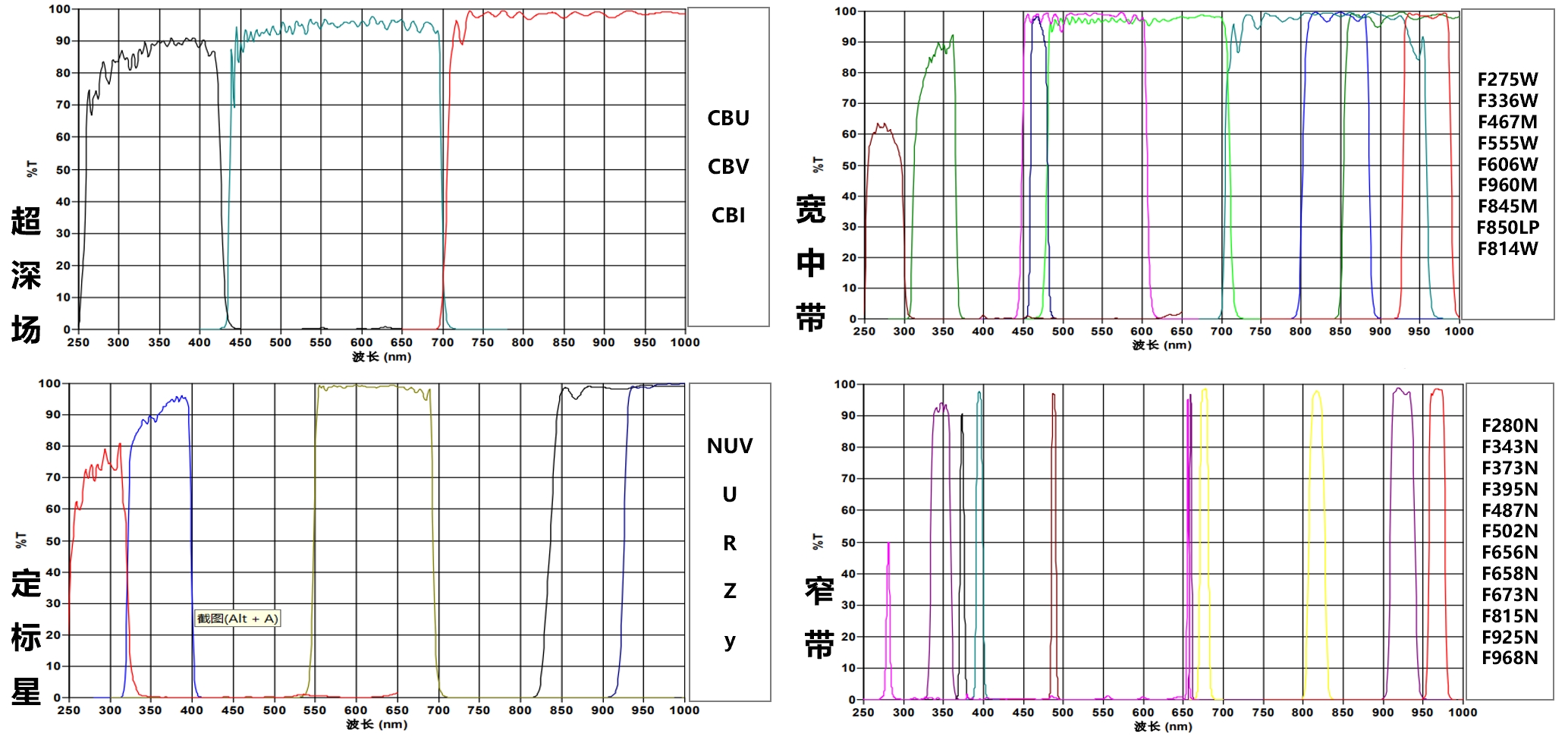}
   \caption{Filter transmission curves of MCI's 30 filters. }
   \label{fig:filters}
   \end{figure}

The MCI is equipped with a total of 30 filters, with 10 filters allocated to each channel. The key parameters of the 30 filters are listed in Tab. \ref{tab:filters}. The filter curves are shown in Fig. \ref{fig:filters}. Among these 10 filters in each channel, 4 filters cover the full field of view, while the remaining 6 filters cover a quarter-FOV area. The CBU filter in the NUV channel, CBV filter in the Blue channel, and CBI filter in the Red channel span nearly the full spectral width of their respective channel. These will be utilized for the MCI extreme-deep field survey covering approximately 300 arcmin$^2$, targeting a limiting magnitude of CBV $\geq$ 30 mag, and CBU and CBI averaging $\geq$ 29 mag via stacking thousands images. Filters such as NUV, u, r, z, and y will support photometric standard stars calibration for the CSST-SC. The remaining filters include narrow-band, medium-band, and broad-band types, primarily dedicated to scientific investigations such as high-redshift galaxies, nearby galaxies, local ionized gas, and objects in our solar system.

For each channel, there are 10 filters installed in one filter wheel, equally distributed in two filter holders. Each filter holder contains 2 filters with the full FOV, and 3 filters with the quarter-FOV. An empty FOV region is kept in each filter holder for the filter change in each channel.

\begin{table}
\begin{center}
\caption[]{Parameters of MCI's 30 filters (ranked by increasing mean wavelengths)}\label{tab:filters}
\begin{tabular}{lccccccccc}
\hline
%
%
Filter Names & FOV & $\lambda_{\rm c}$ & FWHM & $\lambda_{\rm L50}$ & $\lambda_{\rm R50}$ & Tan$_{\rm L50}$ & Tan$_{\rm R50}$ & T50 & {Comments} \\
     &     & (nm)& (nm) & (nm) & (nm) & & & & \\
 (1) & (2) & (3) & (4)  & (5) & (6) & (7) & (8) & (9) & (10) \\
\hline
\multicolumn{10}{c}{MCI Channel 1 (NUV)}  
\\\hline
{\sf F275W} & F & 274.5 & 46.0 & 251.5 & 297.5 & 0.020 & 0.024 & 59.7\% & HST WFC3-F275W\\
{\sf F280N} & P & 279.3 & 3.0 & 277.8 & 280.8 & 0.010 & 0.010 & 62.6\% & MgII, HST WFC3-F280N\\
{\sf NUV} & P & 286.0 & 68.0 & 252.0 & 320.0 & 0.096 & 0.045 & 68.1\% & Calibration \\
{\sf WU} & P & 317.0 & 128.0 & 253.0 & 381.0 & 0.090 & 0.022 & 78.1\% & Calibration \\
{\sf F336W} & F & 338.5 & 53.0 & 312.0 & 365.0 & 0.060 & 0.016 & 81.1\% & HST WFC3-F336W\\
{\sf CBU} & F & 342.5 & 167.0 & 259.0 & 426.0 & 0.050 & 0.056 & 84.2\% & XDF\\
{\sf F343N} & P & 345.5 & 27.0 & 332.0 & 359.0 & 0.022 & 0.016 & 84.9\% & [Ne V], HST WFC3-F343N\\
{\sf u} & P & 360.5 & 77.0 & 322.0 & 399.0 & 0.027 & 0.018 & 89.4\% & Calibration\\
{\sf F373N} & F & 372.9 & 5.4 & 370.2 & 375.6 & 0.007 & 0.007 & 87.3\% & [O II], HST WFC3-F373N\\
{\sf F395N} & P & 395.0 & 8.4 & 390.8 & 399.2 & 0.008 & 0.012 & 86.0\% & v', HST WFC3-F395N\\
\hline
\multicolumn{10}{c}{MCI Channel 2 (Blue)} 
\\\hline 
{\sf F467M} & P & 469.5 & 21.0 & 459.0 & 480.0 & 0.009 & 0.015 & 90.1\% & HST WFC3-F467M \\
{\sf F487N} & P & 486.7 & 5.8 & 483.8 & 489.6 & 0.004 & 0.005 & 89.6\% & H$\beta$, HST WFC3-F487N \\
{\sf F502N} & F & 501.2 & 7.0 & 497.7 & 504.7 & 0.004 & 0.006 & 88.1\% & [O III], HST WFC3-F502N\\
{\sf F555W} & P & 527.5 & 157.0 & 449.0 & 606.0 & 0.019 & 0.014 & 97.1\% & HST WFc3-F555W\\
{\sf CBV} & F & 567.5 & 261.0 & 437.0 & 698.0 & 0.013 & 0.012 & 94.2\% & XDF \\
{\sf F606W} & F & 595.0 & 230.0 & 480.0 & 710.0 & 0.018 & 0.014 & 96.2\% & HST WFC3-F606W \\
{\sf r} & P & 620.5 & 145.0 & 548.0 & 693.0 & 0.018 & 0.014 & 97.3\% & Calibration\\
{\sf F656N} & P & 656.2 & 1.7 & 655.3 & 657.0 & 0.002 & 0.002 & 83.5\% & H$\alpha$, HST WFC3-F656N\\
{\sf F658N} & F & 658.1 & 8.2 & 654.0 & 662.0 & 0.003 & 0.004 & 91.6\% & H$\alpha$-wide, HST WFC3-F658N\\
{\sf F673N} & P & 676.5 & 11.6 & 670.7 & 682.3 & 0.006 & 0.006 & 91.0\% & H$\alpha$-off, HST WFC3-F673N\\
\hline
\multicolumn{10}{c}{MCI Channel 3 (Red)} 
\\\hline 
{\sf F815N} & F & 815.0 & 18.0 & 806.0 & 824.0 & 0.007 & 0.007 & 92.3\% & LAEs at z$\sim$5.7 \\
{\sf F814W} & F & 834.5 & 251.0 & 709.0 & 960.0 & 0.012 & 0.019 & 97.4\% & HST WFC3-F814W\\
{\sf F845M} & P & 845.0 & 86.0 & 802.0 & 888.0 & 0.012 & 0.013 & 95.6\% & LBGs at z$\sim$5.6--6.3\\
{\sf CBI} & F & 852.5 & 293.0 & 706.0 & 999.0 & 0.020 & - & 97.6\% & XDF\\
{\sf F925N} & P & 924.0 & 30.0 & 909.0 & 939.0 & 0.006 & 0.011 & 91.0\% & LAEs at z$\sim$6.6 \\
{\sf z} & P & 924.0 & 158.0 & 845.0** & 1000.0 & 0.012 & 0.017 & 97.3\% & Calibration\\
{\sf F960M} & P & 958.5 & 59.0 & 929.0 & 988.0 & 0.006 & 0.007 & 96.8\% & LBGs at z$\sim$6.7--7.1\\
{\sf F968N} & P & 968.0 & 20.0 & 958.0 & 978.0 & 0.004 & 0.006 & 92.8\% & LAEs at z$\sim$7.0\\
{\sf F850LP} & F & 976.0 & 244.0 & 854.0 & 1098.0 & 0.012 & - & 97.7\% & HST WFC3-F850LP\\
{\sf y} & P & 1012.0 & 172.0 & 926.0 & 1098.0 & 0.012 & - & 98.2\% & Calibration\\
\hline 
\end{tabular}
\end{center}
\raggedright {Notes: Column (1): Name of the filter. Column (2): FOV covered by the filter: "F" for full FOV (7\arcmin.5 $\times$ 7\arcmin.5), and "P" for one quarter FOV (3\arcmin.75 $\times$ 3\arcmin.75). Column (3): The central wavelength ($\lambda_{\rm c}$) of the filter. Column (4): The full-width at half-maximum (FWHM) of the filter. Column (5): The blue-side wavelength at the 50\% of the peak throughput ($\lambda_{\rm L50}$). Column (6): The red-side wavelength at the 50\% of the peak throughput ($\lambda_{\rm R50}$). Columns (7) and (8): The steepness of the filter at $\lambda_{\rm L50}$ ($\lambda_{\rm R50}$), defined as the slope of the filter transmittance at the increase of $\lambda_{\rm L50}$ (or the decrease of $\lambda_{\rm R50}$) extends to the wavelength difference of $\delta\lambda$ at the transmittance of 0\% and 100\%, divided by the wavelength at the point, i.e., Tan$_{\rm L50}$ =$\delta\lambda/\lambda_{\rm L50}$ (or Tan$_{\rm R50}$ = $\delta\lambda/\lambda_{\rm R50}$). Column (9): Average throughput (T50) in the wavelength range of $\lambda_{\rm L50}$ and  $\lambda_{\rm R50}$. Column (10): Comments on the filter usage or similar HST filters. }
\end{table}

%

\subsection{CCD}
\label{sect:Design:MCICCDs}
\begin{figure}
   \centering
   \includegraphics[width=\textwidth, angle=0]{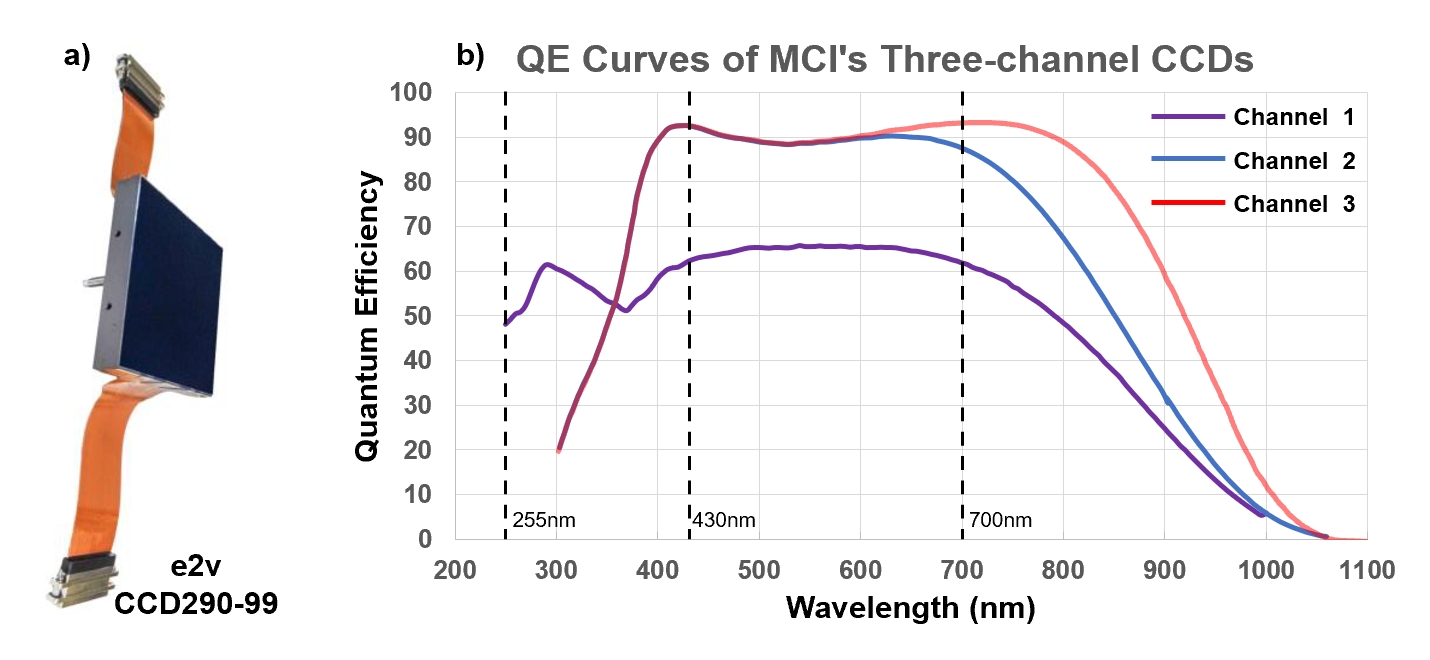}
   \caption{a): e2v CCD290-99; b): Quantum efficiency curves of MCI's CCDs in three channels. }
   \label{fig:QEs}
   \end{figure}

The CCD detector chosen for MCI is the e2v CCD290-99 sensor optimized for each channel's sensitivity. The sensor has an imaging area of 9216 × 9232 pixels with registers at both top and bottom, each with eight outputs (see Fig. \ref{fig:QEs}-a) for short read-out time. The pixel size is 10 $\mu$m square. The operating temperature of the CCD is at 173K (-100 $^\circ$C). We require the dark current value of $<$0.02 e$^-$/pixel/s. To balance the readout speed and noise, we set the readout time of 40 sec and the required readout noise $<$ 5 e$^-$.

 The CCD chosen for each channel is optimized for the total throughput in the corresponding channel. A CCD with UV enhanced coating is equipped in the NUV channel. A CCD with standard silicon and multi-II coating is equipped in the blue channel. A CCD with deep depletion and multi-II coating is equipped in the red channel.  The spectral response curve (or the quantum efficiency curve) for each channel's CCD is shown in Fig. \ref{fig:QEs}.  

%

\subsection{Sensitivity and Exposure Time Calculator}

\begin{figure}
   \centering
   \includegraphics[width=\textwidth, angle=0]{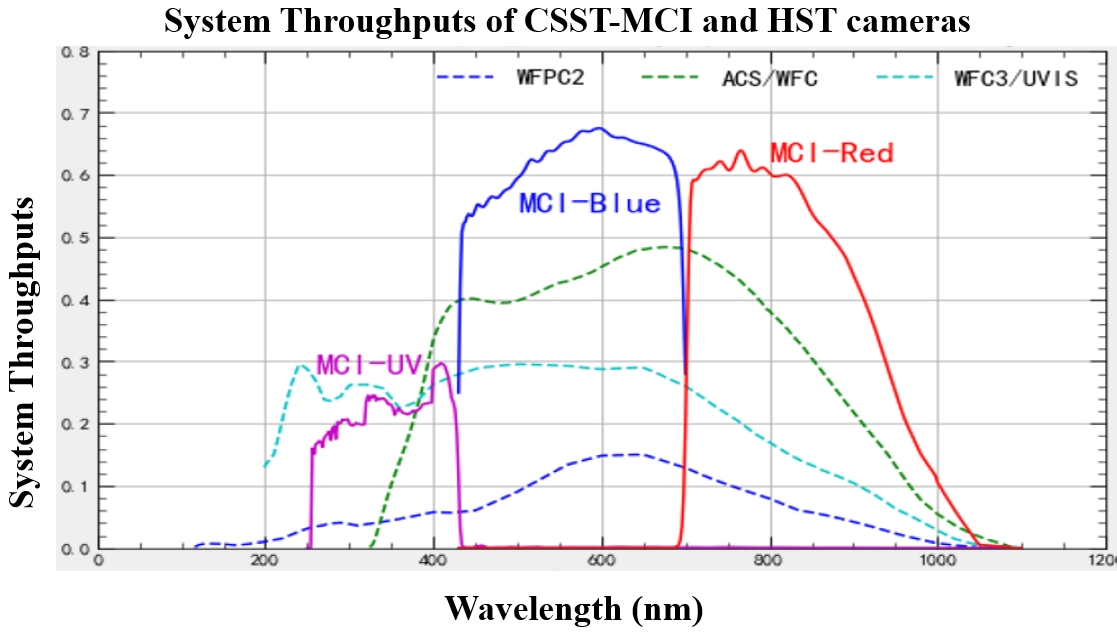}
   \caption{System throughputs of the CSST-MCI's three channels and the HST cameras.}
   \label{fig:throughputs}
   \end{figure}

Currently, the total throughput of the CSST-MCI is not measured yet but estimated by multiplying the throughput of the primary optical system and that of the MCI instrument.
{In each channel of MCI the throughput is the multiplication of reflecting coefficients on that light path. }
Because of the extra optical system inside MCI, the total throughput of CSST-MCI is about 3/4 of that of CSST-SC, averaged over the whole wavelength coverage. {Note that this is a preliminary estimate prior to the full system measurement.} Despite that, the system throughput of MCI is comparable to that of the HST cameras, which are shown in Fig. \ref{fig:throughputs}. Direct measurements of the total throughputs will be done with the flight model in about one year. 

We have developed an exposure-time calculator (ETC) for MCI which is located on the Chinese Virtual Observatory website 
\footnote{MCI ETC link: \url{https://nadc.china-vo.org/csst-bp/etc-ms/etc-mci.jsp}} . 
The MCI ETC is designed to provide quantitative estimates of exposure time, signal-to-noise ratio (SNR), and limiting magnitude for specific observational targets, taking into account the optical performance of the CSST telescope and the hardware properties of the MCI module. 

The ETC supports both point sources and extended sources. Extended sources are modeled with parameterized profiles, including uniform disk, Sersic, and Gaussian morphologies. For spectral inputs, the tool accommodates stellar model spectra \citep{2003IAUS..210P.A20C}, representative galaxy spectra, blackbody spectra, as well as user-defined spectra. The sky background model follows the background curves adopted in the HST ACS exposure time calculator \citep{2012acsi.book...12U}. Over the CSST wavelength range, the dominant contribution of the sky comes from the zodiacal light and the airglow. In addition, an extinction model is applied, allowing extinction to be estimated either from the Galactic dust map \citep{1998ApJ...500..525S} in a specified direction or from user-defined extinction coefficients.

Here in Fig. \ref{fig:depth}, we present the 5-$\sigma$ limiting magnitudes of a point source with a flat spectrum for all these 30 filters with exposure times of 30 sec, 300 sec, and 6x300 sec calculated with the ETC under typical observing conditions. By combining these inputs with the instrument characteristics of the CSST MCI, the ETC provides a practical tool for evaluating the detectability and observational feasibility of a wide range of astrophysical targets. 
The ETC will be updated {after flight-model throughput measurements and on-orbit calibration.}

%

   \begin{figure}
   \centering
   \includegraphics[width=\textwidth, angle=0]{Depth.png}
   \caption{5-$\sigma$ limiting magnitudes of a point source with a flat spectrum for all these 30 filters with exposure times of 30 sec (top panel), 300 sec (middle panel), and 6x300 sec (bottom panel) calculated with the ETC.}
   \label{fig:depth}
   \end{figure}

\section{Ground Tests}
\label{sect:test}

Here we briefly introduce the detector noise and the image quality obtained from the ground testing experiments taken in the past two years. The detector noise and the image quality are among the most important features to evaluate the performance of the MCI. 
Detailed results of the ground tests will be released in a series of technical papers later. Furthermore, there will be a final round of ground tests in the next year for the flight model of the CSST and the MCI, which will cover most of the features required for the pre-launch calibrations.

\subsection{Detector Noise}
The dark current and the readout noise are the main noises generated by the CCD detector. For each channel's CCD, the design requirement of the dark current is $<$0.02 e$^-$/pixel/s, and that of the readout noise is $<$5e$^-$. In the ground testing, we measured median values of dark current of $<$0.001 e$^-$/pixel/s at 173K for CCDs in the Blue and Red channels, and of $<$0.005 e$^-$/pixel/s at 173K for the CCD in the NUV channel. The measured readout noise for each channel's CCD is $<$4 e$^-$ and meanwhile the full well depth is over 70k e$^-$ in each channel. They all indicate good detector performance and satisfy the design requirements.

\subsection{Image Quality}
The required dynamic image quality of the CSST-MCI is  R$_\textrm{EE80}\leq$ 0\arcsec.18, where R$_\textrm{EE80}$ is the radius of a point source that encircles 80\% of the energy. This image quality is a combined effect of both the main optical system and the MCI.
For the MCI alone, according to the tolerance decomposition of the image quality, the required image quality is R$_\textrm{EE80}\leq$ 0\arcsec.12, or $\leq$ 2.4 pixels. 

The image quality for the MCI alone was tested at SITP with a pinhole plate illuminated by parallel light. The plate has a 13x13 pinhole array covering the whole FOV of the MCI. After the basic image data reduction and the peak finding, we fit a 2-d Gaussian to each individual pinhole. The average R$_\textrm{EE80}$ and ellipticity for MCI's NUV, Blue and Red channels are [1.5$\pm$0.2, 1.6$\pm$0.2 and 1.6$\pm$0.1] pixels and [0.021, 0.011, 0.0001], respectively. This indicates that the image quality of the MCI alone is quite good. 

The test of image quality of CSST-MCI, when assembled with the main optical system, was taken at the Changchun Institute of Optics, Fine Mechanise and Physics (CIOMP, CAS) with a simulated point light source. 
It was lighted at 9 positions in the FOV of the MCI. We measured that all the 9 points in each channel have R$_\textrm{EE80}<$ 0\arcsec.18 and show a very sharp shape, which meets the image quality requirement. Note that the measured R$_\textrm{EE80}$ value includes the uncertainty caused by simulating the point light source. Therefore, the final image quality for the CSST-MCI is quite good as well.

\section{Suggested Observing Strategies of the MCI}
\label{sect:obs}
In this section we suggest some observing strategies of the MCI.
As shown in Tab. \ref{tab:comparison}, unlike HST or JWST cameras, only the MCI can observe the same target in three channels simultaneously.
It can not only improve the observing efficiency, but also provide a simultaneous color image of the same target, which is very important for short time-series analysis on special targets such as transiting exoplanets \citep{2025ChA&A..49..436D}, rotating asteroids and variable comets in our solar system. 

The MCI's three-channel simultaneous imaging can also improve the performance of space-based deep narrowband imaging and NUV imaging. After checking the available HST narrowband imaging surveys, we found that most HST narrowband observations focused on local universe objects, e.g., local ultra-faint dwarf galaxies \citep{Fu2022}, young stellar objects \citep{Ferreras2009}, and supernova remnants (i.e., SN1987A, \citealt{Larsson2013}) in the Large Magellanic Cloud. There are quite few deep HST narrowband imaging data available except for our HDH$\alpha$ project \citep{Zhu2024}. With the HDH$\alpha$ project, we have found that the space-based single channel narrowband imaging would suffer from cosmic rays. Besides, there were quite few bright objects in the narrowband image for good image alignments, which is essential for deep exposures through the image stacking. However, 
with simultaneous imaging in three channels, we can choose narrowband imaging in one channel and broadband imaging in another two channels. The synchronized broadband images can help the narrowband images to get rid of the cosmic rays and to have good alignments in stacking for a deep image. In the space-based NUV imaging, the same problem exists as in the narrowband imaging,  i.e., a lot of cosmic rays and quite few bright objects. With MCI's three-channel simultaneous imaging, we can have deep NUV images as well.

\begin{figure}
   \centering
   \includegraphics[width=\textwidth, angle=0]{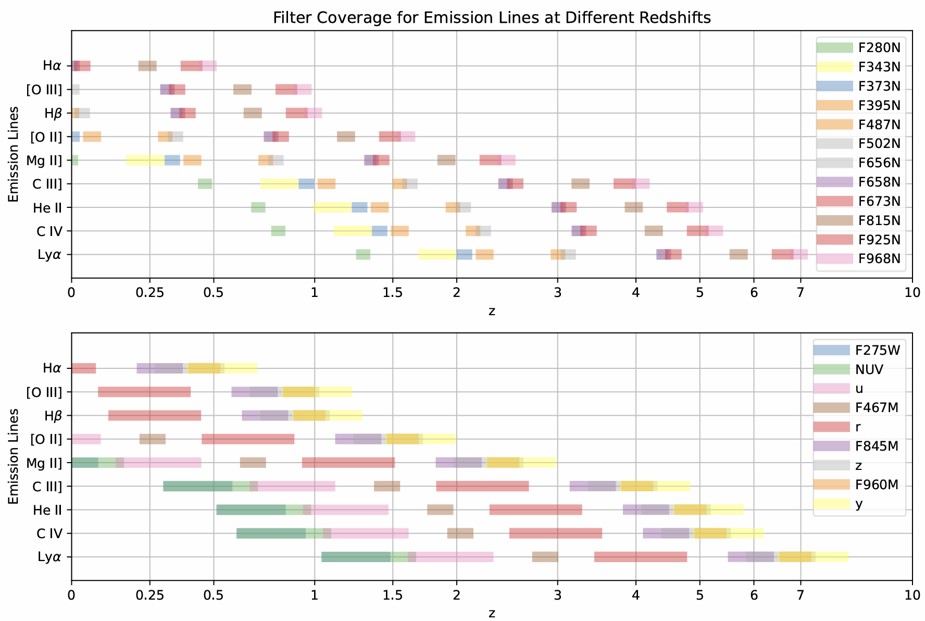}
   \caption{Covered emission lines as a function of redshift probed by narrowband and medium band filters of the MCI.}
   \label{fig:MCIzlines}
   \end{figure}

The rich filter sets of the MCI can help cover the emission lines such as H$\alpha$, [O\,{\sc iii}], H$\beta$, [O\,{\sc ii}] and Mg\,{\sc ii}] in the local Universe and Ly$\alpha$, C\,{\sc iv}, He\,{\sc ii}], and C\,{\sc iii}] at high redshifts, which are shown in Fig. \ref{fig:MCIzlines}. The narrowband filters onboard the MCI can help search Ly$\alpha$ emitters at redshifts in the range of 1.3$\lesssim z \lesssim$7.0.


The CSST-MCI has both the staring mode and the scanning mode, depending on the science objective. There are also dithering patterns in the staring mode to help get rid of the hot and bad pixels in the CCDs. 

Furthermore, the MCI can do parallel  observing in a staring mode with the IFS and the CPI-C, since all these three instruments share a same focal plane. In that focal plane, the IFS FOV is located at the center position of [0$\arcmin$.0,0$\arcmin$.0]. The centers of the MCI FOV and the IFS FOV have a fixed separation of about [0$\arcmin$.0, 25$\arcmin$.2]. The centers of the MCI FOV and the CPI-C FOV have a fixed separation of about [-18$\arcmin$.0, 50$\arcmin$.4]. When the CSST is taking long exposures with the IFS or the CPI-C, the MCI can do parallel deep imaging in the nearby sky.

\section{Summary}
\label{sect:sum}
As one of the five first-generation instruments onboard the CSST, the MCI offers the capability of three-channel simultaneous imaging, with 30 filter covering a $\sim$7.5 x 7.5 arcmin$^2$ FOV. The 9kx9k e2v CCD in each channel of the MCI has a dark current value of $<$0.02 e$^-$/pixel/s and a readout noise $<$5 e$^-$. The total throughput of the MCI is comparable to that of the HST's latest cameras. The MCI has an angular resolution of 0\arcsec.05 per pixel, and a dynamic image quality of R$_\textrm{EE80}<$ 0\arcsec.18. We have developed an ETC for evaluating the detectability and observational feasibility for the CSST-MCI. 

The CSST-MCI's three-channel simultaneous imaging can not only improve the observing efficiency, but also provide real-time colors essential for short time-series analysis and is specially suitable for deep NUV and narrowband imaging. Besides the staring mode, MCI can also do the scanning mode which can help study the moving objects in our Solar system. Finally, the MCI can do parallel observing with the IFS and the CPI-C.

\begin{acknowledgements}
We would like to extend our sincere gratitude to Academician Chen Jiansheng and Professor Hu Jingyao of the National Astronomical Observatories for proposing and promoting the survey project based on the large space-based optical platform (now known as the CSST). We are also deeply grateful to Academician Gu Yidong of the Technology and Engineering Center for Space Utilization for his tremendous dedication to the CSST and especially to the MCI. Additionally, we wish to express our appreciation to the China Manned Space Engineering program and the Space Application System for their strong support. We would like to express our gratitude to the experts and scholars from both China and abroad for their strong support of this project. We thank the anonymous referee for the suggestions and comments which significantly improve our paper.
This work is supported by the China Manned Space Program, and partly supported by grant no. CMS-CSST-2025-A18.
\end{acknowledgements}




\label{lastpage}

\end{document}